\documentstyle[preprint,prl,aps]{revtex}
\newcommand{\be}{\begin{equation}}
\newcommand{\ee}{\end{equation}} 
\newcommand{\bea}{\begin{eqnarray}}
\newcommand{\eea}{\end{eqnarray}}
\newcommand{\eid}{\stackrel{\mathrm{d}}{=}}
\def \Re{{\rm I\kern -1.6pt{\rm R}}}
\def \Expect{{\rm I\kern -1.6pt{\rm E}}}
\tightenlines

\begin{document}

\draft

\title{Better Nonlinear Models from Noisy Data: Attractors with 
Maximum Likelihood}

\author{Patrick E. McSharry\cite{byline1} and 
Leonard A. Smith\cite{byline2}}

\address{Mathematical Institute, University of Oxford, Oxford OX1 3LB.}

\date{\today}

\maketitle

\begin{abstract}
A new approach to nonlinear modelling is presented  which, by
incorporating the global behaviour 
of the model, lifts shortcomings of both least 
squares and total least squares parameter estimates. 
Although ubiquitous in practice, a least squares approach is fundamentally 
flawed in that it assumes independent, normally distributed (IND) 
forecast errors: nonlinear models will not yield IND errors even if the 
noise is IND. 
A new cost function is obtained via the maximum likelihood principle;
superior results are illustrated both for small data sets and infinitely long
data streams. 
\end{abstract}
\pacs{PACS numbers: 05.45.-a, 05.45.Tp, 02.60.-x}

\narrowtext
A nonlinear model must be tuned via parameter estimation, ideally  
forcing it to mimic the observations. 
Typically, tuning aims for parameters which yield the least 
squared error \cite{bjorck96,grasset91,jaeger96} (or total least squared 
error \cite{vanhuffel91,kostelich92}) between the one-step forecasts
and the data. After proving that even for the simplest nonlinear models, 
both least squares and total least squares systematically reject the correct 
parameter values (i.e. those that generated the data), a new and more 
robust method is derived which incorporates the global dynamics of the 
model (and hence its attractor). 
Failure to recognise the effects of imperfect observations
will lead to biased parameter estimates, whereas an inability to reflect 
the data indicates model error. 
The present Letter focuses on the first issue: using the fact that one can 
always estimate the probability density function (PDF) of the model-state
variable for different values of the unknown parameters, 
the maximum likelihood principle is employed to derive a new cost function 
which incorporates this information. 
This global approach has the potential to outperform all one-step (or
few-step) methods, whether they are based on least squares criteria or 
some future improvement. 
Note that even with an infinite amount 
of data, the optimal least squares solution is simply incorrect; 
see Fig. \ref{fig1} and the discussion 
below. The new cost function is shown to yield results consistent with the 
correct answer even for relatively small data sets and large noise levels in 
a variety of chaotic systems. It is applicable to high dimensional systems 
and may also be applied to nonlinear stochastic systems.

Suppose the evolution of a system's state variable, 
\mbox{${\bf x}_i \in \Re^m$}, is 
governed by the map 
\be
{\bf x}_{i+1} = {\bf F}({\bf x}_i, {\bf a}),
\label{e:functrel}
\ee
where the model's parameters are contained in the vector 
\mbox{${\bf a} \in \Re^l$}. For $m$ = 1, the system state $x_i$ is
a scalar; assuming additive measurement noise $\eta_i$ yields
observations $s_i = x_i + \eta_i$. 
In a noise free setting (i.e. $\eta_i = 0 \ \forall ~i$), $l$ + 1  
sequential measurements $s_i,s_{i+1},...,s_{i+l}$  
would, in general, be sufficient to determine ${\bf a}$. 
With noise, the PDF of both the measurement noise and the
model-state variables are required to estimate ${\bf a}$ properly.

Given the correct model structure $F({\bf x}, {\bf a})$ and data 
generated by particular parameters ${\bf a}_0$ 
(the `true' parameter values), a cost function is often used to obtain 
an estimate of the unknown parameters ${\bf a}_0$. 
Ideally, this estimate converges to ${\bf a}_0$ in the limit of an infinite 
number of observations.
Complications of nonlinearity combined with the
randomness of unavoidable measurement errors suggests a likelihood analysis 
for parameter estimation. 
Indeed, both least squares and total least squares 
are special cases of the likelihood method.

The one-step {\it least squares} (LS) estimate, ${\bf a}_{LS}$,
is the value of ${\bf a}$ 
which minimises the least squares cost function 
\be 
C_{\rm LS}({\bf a}) = \sum_{i=1}^{N-1} E_i^2,
\label{e:lscf}
\ee
where \mbox{$E_i = s_{i+1} - F(s_i,{\bf a})$}, the one-step 
prediction error. 
Figure \ref{fig1} shows the failure of this well known  
technique when applied to the well known Logistic Map \cite{may76}:
$a_0$ = 2, yet $a_{LS} < 2$ at all nonzero noise levels,
even for an infinite data set.
To see this, first recall this one-dimensional map $F(x,a) = 1 - a x^2$.
Writing the observed prediction error, \mbox{$s_{i+1} - F(s_i,a)$}, 
explicitly in terms of the underlying system state and a realisation of 
the noise process yields 
\mbox{$E_i = \eta_{i+1} - a_0x_i^2 + a(x_i + \eta_i)^2.$}
When the least squares cost function (\ref{e:lscf}) 
is a minima \mbox{$\sum_{i=1}^N \left[ 
E_i \partial E_i / \partial a \right] = 0$}, and in
the limit of an infinitely long  data set this sum converges to an 
integral over $x$ taken with respect to the system's invariant 
measure, $\mu(x,a)$. 
For $a = 2$ in the Logistic map, \mbox{$\mu(x,2)=1/\pi \sqrt{1 - x^2}$} 
for $-1 \leq x \leq 1$ and zero otherwise.   
Thus $\langle x^2 \rangle$ = 1/2 and $\langle x^4 \rangle$ = 3/8. 
Since $\langle x^n \rangle$ = 0 for odd $n$ 
($\mu(x,2)$ is an even function), if the distribution of the noise process 
is also even, then
\be
a_{LS} = \left( \frac{4 \langle \eta^2 \rangle + 3}
{8 \langle \eta^4 \rangle + 24 \langle \eta^2 \rangle + 3} \right) a_0.
\label{e:aopt}
\ee 
Equation (\ref{e:aopt}) yields the parameter estimate corresponding to 
an infinitely long data set as a function of noise level.
For uniformly distributed noise \mbox{[i.e. $\eta \eid
U(-\epsilon,\epsilon)$],}    
\mbox{$\langle \eta^{n} \rangle = \epsilon^{n}/(n+1)$} for even $n$, 
while for normally distributed noise  
[i.e. \mbox{$\eta \eid N(0,\epsilon^2)$}],   
\mbox{$\langle \eta^2 \rangle = \epsilon^2$} and  
\mbox{$\langle \eta^4 \rangle = 3 \epsilon^4$}, 
The noise level is defined as   
\mbox{$\sigma_{noise}/\sigma_{signal}$}, where $\sigma^2_{noise}$ and 
$\sigma^2_{signal}$ are the variances of the noise and the signal, 
respectively; for the uniform case $\sigma^2_{noise} =
\epsilon^2/3$, whereas $\sigma^2_{noise} = \epsilon^2$ for the normal case. 
Despite having a complete knowledge of the measurement process and data of 
infinite duration, the parameter estimates are biased.

Let $(x_i,x_{i+1})$ denote a successive pair of system states  
corresponding to observations $(s_i,s_{i+1})$ where
\be
s_i = x_i + \eta_i, \ \ \ 
\eta_i \eid N(0,\epsilon^2).
\label{e:measproc}
\ee 
The system variables $(x_i,x_{i+1})$ are sometimes called {\it latent} 
variables, due to the fact that they cannot be 
measured directly \cite{casella90}. 
They are fixed yet unknown, and therefore all 
probabilities ought to be conditional on the $x_i$. 
In general, the probability of observing the pair $(s_i,s_{i+1})$ 
depends on the model parameters ${\bf a}$, the system variable   
$x_i$ at time $i$, its image $F(x_i,{\bf a})$, and the measurement process. 
Specifically  
\be
P(s_i,s_{i+1} | {\bf a}, x) = \frac{1}{2\pi \epsilon^2}
\exp \left( - \frac{d_i^2}{2 \epsilon^2}  \right),
\label{e:pxiyigaxii}
\ee
where 
\be
d_i^2(s_i,s_{i+1},x,{\bf a}) 
= (s_i - x)^2 + (s_{i+1} - F(x,{\bf a}))^2.
\label{e:di}
\ee
{\it Assuming} the $s_i$ and $s_{i+1}$ are independent, the probability of 
observing a sequence of $N-1$ pairs, 
\mbox{${\bf S} = \{(s_i,s_{i+1})\}_{i=1}^{N-1}$}, 
corresponding with a particular set of model-states 
$\tilde{\bf X} = \{\tilde{x}_i\}_{i=1}^{N-1}$ 
is given by the joint PDF:
\be
P({\bf S} | {\bf a}, \tilde{\bf X}) 
= \prod_{i=1}^{N-1} P(s_i,s_{i+1} | {\bf a}, \tilde{x}_i).
\label{e:pxygaxi}
\ee

Identifying the {\it likelihood} of parameters ${\bf a}$ 
generating the data ${\bf S}$
with the probability of observing data ${\bf S}$ given that the 
model has parameters ${\bf a}$ (see \cite{press92}) yields  
\be
L({\bf a}, \tilde{\bf X} | {\bf S}) 
= P({\bf S} | {\bf a}, \tilde{\bf X}),
\label{e:laxigxyp}
\ee
where the conditional status of the model-state variables is explicit.   
Substituting (\ref{e:pxiyigaxii}) and (\ref{e:pxygaxi}) in
(\ref{e:laxigxyp}) yields  
\be
L({\bf a}, \tilde{\bf X} | {\bf S}) 
= \frac{1}{(2\pi \epsilon^2)^{N-1}}
\exp \left( - \frac{1}{2 \epsilon^2} 
\sum_{i=1}^{N-1} d_i^2 \right).
\label{e:laxigxy}
\ee 
$L({\bf a}, \tilde{\bf X} | {\bf S})$ depends  
on the PDF of the likely model-states and the 
parameters which maximise (\ref{e:laxigxy}) will vary with
the assumptions made regarding this distribution. 
These assumptions are paramount to this letter;
ignoring information from the distribution will lead to 
{\it total least squares} (TLS), 
whereas requiring consistency between the data and 
the PDF of the model-state variables yields the new cost function below. 
Casella and Berger \cite{casella90} compare these assumptions for 
linear systems.

Ignoring the PDF of the model-state variables, total least squares 
resolves the dependence on $\tilde{x}_i$ by substituting any
values $\tilde{x}_i$ which maximise $L({\bf a} | {\bf S})$, 
that is:
\be
L({\bf a} | {\bf S}) 
= \frac{1}{(2\pi \epsilon^2)^{N-1}}
\exp \left( - \frac{1}{2 \epsilon^2} 
\sum_{i=1}^{N-1} \min_{x \in \Re} d_i^2 \right).
\label{e:likef}
\ee 
The maximum of $L({\bf a},{\bf S})$ then corresponds to the minimum of the 
associated TLS cost function 
\be
C_{\rm TLS}({\bf a}) 
= \sum_{i=1}^{N-1} \min_{x \in \Re} d_i^2.
\label{e:tlscf}
\ee 
Thus, while the least squares cost function (\ref{e:lscf}) 
minimises the squared vertical
distances $d_i^2 = \left[ s_{i+1} - F(s_i,{\bf a}) \right]^2$,  
the TLS solution minimises the squared perpendicular 
distances (\ref{e:di}) between the measured point $(s_i,s_{i+1})$ and a 
point on the hyper-surface $(\tilde{x}, F(\tilde{x}, {\bf a}))$. 
No restrictions are placed on the values $\tilde{x}_i, i=1,N$: 
the $\tilde{x}_i$ are not a trajectory of $F(x,{\bf a})$, nor do
they reflect $\mu(x,{\bf a})$. 
Using the particular $\tilde{x}_i$ which minimises each $d_i^2$ reflects 
the decision to ignore any knowledge of the PDF of the model-state variables. 
But, given that the model is always in hand the PDF of the model, 
$\mu(x, {\bf a})$, is always obtainable (not that of the system, but of 
the model).
This additional information may be incorporated by integrating the 
dependence on $x_i$ out of the likelihood function in (\ref{e:laxigxy}),
yielding 
\be 
L({\bf a} | {\bf S}) 
= \prod_{i=1}^{N-1} \int_x P( s_i, s_{i+1} | {\bf a}, x) d \mu(x, {\bf a}),
\label{e:likes}
\ee
and associated {\it maximum likelihood} (ML) cost function:
\be
C_{\rm ML}({\bf a}) = - \sum_{i=1}^{N-1} \log 
\int_x \exp \left(- \frac{d_i^2}{2 \epsilon^2} \right)
d \mu(x, {\bf a}).
\label{e:mlcf}
\ee
Equation (\ref{e:mlcf}) is the main result of this Letter;   
it is superior to both $C_{\rm LS}$ and $C_{\rm TLS}$. 
In practice, the integral in (\ref{e:mlcf}) is usually
replaced by a sum over a 
model trajectory whose length, $\tau >> N$, 
is limited only by computational constraints. Thus  
\bea
C_{\rm ML}({\bf a}) &\approx& - \sum_{i=1}^{N-1} \log \Bigg( 
\sum_{k=1}^{\tau} \exp \bigg\{ - \frac{1}{2 \epsilon^2} 
\Big[ (s_i - x_k)^2 
\nonumber \\
&&+ 
(s_{i+1} - x_{k+1})^2 \Big] 
\bigg\} \Bigg),
\label{e:mlcfsum}
\eea
where $x_k = F^k(x_0,{\bf a})$ and $x_0$ is any post transient value 
$F(x, {\bf a})$ in the relevant basis of attraction. 
For fixed $N$ this can be computed efficiently \cite{schreiber95}.
As $\tau \to \infty$, the particular value of $x_0$ is irrelevant since the 
sum is dominated by those values of $k$ for which $s_i \approx x_k$ and 
$s_{i+1} \approx x_{k+1}$, the particular values of $k$ being irrelevant.

The Logistic map's invariant measure varies 
drastically for different values of $a$. 
The bifurcation diagram (see Fig. \ref{fig2}a) 
illustrates this behaviour in the range $1.5 \leq a \leq 2$. 
With $a_0$ = 2 and a noise level of $0.19$, $a_{LS}$ = 1.7 
(see Fig. \ref{fig1});
note from Fig. \ref{fig2}a that this corresponds to a model with 
a period three orbit.
Obviously, values of $a$ corresponding with 
periodic windows are unlikely to be responsible for a data set with 
wildly aperiodic behaviour.
Equation (\ref{e:mlcfsum}) allows this visually
obvious result to be included implicitly into a cost function.

The TLS cost function for the Logistic map may be obtained 
analytically by solving the cubic equation 
\be
\frac{\partial d_i^2}{\partial x_i} 
= 4 a^2 x_i^3 + [2 + 4(s_{i+1} - 1)a] x_i - 2s_i = 0,
\ee
and taking the root satisfying $\partial^2 d_i^2 / \partial x_i^2 > 0$. 
Results for three different cost functions are shown in 
Fig. \ref{fig2}. 
While the TLS cost function is much better than that of LS, the ML 
cost function is better still for all cases considered. 
Results are shown for data sets $N$ = 100;
for larger $N$ the TLS solution improves, but 
in all cases tested the spread of the 
distribution remains smaller for the ML estimate. 
The simplicity of the Logistic map make it a weak test case and 
motivates further trials.

The Moran-Ricker map \cite{maymr76} has a functional form
$F(x,a) = x \exp [ a(1- x)]$; it's invariant measure (not shown)
allows larger values of $x$ with 
larger parameter values $a$.
Figure \ref{fig3} illustrates the results for each cost functions 
where $a_0$ = 3.7 and $N$ = 100. 
The ML cost function consistently
yields the best estimates for all noise levels considered. 
While it is common to claim an algorithm generalises to higher dimensional 
cases, this algorithm is easily generalised to $m > 1$ by substituting 
the term $\left[ || {\bf s}_i - {\bf H}({\bf x}_k) || \right]$ 
for the term $\left[ (s_i - x_k)^2 \right]$ in equation (\ref{e:mlcfsum}). 
Here the function ${\bf H}$ projects the system state vector ${\bf x}$ 
into the space 
of observations ${\bf s}$. In delay coordinates, this corresponds to taking the
`last' component of ${\bf x}_k$. Fig. \ref{fig4} shows results for the 
two-dimensional H\'{e}non map \cite{henon76} in delay coordinates,
$F(x_i,x_{i-1}) = 1 - a x_i^2 + b x_{i-1}$, where $a_0 = 1.4$ and 
$b_0 = 0.3$. 
While the ML cost function surface is more highly structured due to 
sensitivity to the parameters, its minima are in the relevant regions as 
opposed to the smooth but incorrect LS minimum. 
The LS cost function has a biased
minima, while ML is consistent with $(a_0,b_0)$. 
Whether this consistency
is worth the computation depends on the problem at hand.
Certainly the fact that 
ensemble forecasts of chaotic models using the ML estimates will
relax naturally to a distribution consistent with $\mu({\bf x}, {\bf a})$
is of value \cite{shadow}. 
Better estimates of ${\bf a}$ also allow improved long term deterministic 
forecasts. 
The fact that higher dimensional models may require much larger data
sets is a problem of uniqueness under the observations and cannot
be laid at the door of the cost function. In Figure 4, $N=500$ and the 
noise level is 0.05.
Equation (\ref{e:mlcfsum})
also allows an estimate of the magnitude of dynamical noise 
in a stochastic system \cite{shadow} when the shape of the 
distribution of the noise is correctly specified.

Both least squares (\ref{e:lscf}) and total least squares (\ref{e:tlscf}), 
are inferior to the maximum likelihood cost function (\ref{e:mlcf}).
Including the information on the invariant measure of 
the model aids in the pursuit of 
reliable parameter estimates for nonlinear models. 
The weakest link in the derivation of the ML cost function 
is the assumption that the $s_i$ and $s_{i+1}$ are independent, while 
they may be linearly uncorrelated, they cannot be independent. 
Attempts to relax this assumption will be presented in later work; here we 
note that (\ref{e:mlcf}) may be viewed as the first member
of the family of cost functions 
\be
C^{(n)}_{\rm ML}({\bf a}) = \sum_{i=1}^{N-n} 
\int \exp \left(- \frac{{d^{(n)}_i}^2}{2 \epsilon^2} \right)
d \mu(x_i, {\bf a}),
\label{e:mlcfk}
\ee
where 
\be
{d^{(n)}_i}^2 = \sum_{j=0}^n [s_{i+j}-F^j(x_i, {\bf a})]^2.
\ee
For $n > 0$ $C^{(n)}_{\rm ML}({\bf a})$ is a multi-step
cost function \cite{jaeger96,gilmour} moderating the  
assumption that the $s_i$ are independent,  
the aim being to find parameter values which both have the
correct PDF and shadow the observations \cite{shadow,gilmour}. 
The aultimate $n=N-1$ multi-step least squares approach, solving
simultaneously for $x_0$ and $a$, may prove intractable even for $N=500$ 
(see \cite{berliner91}). 
Future work will also focus upon the choice of optimal model order in 
local polynomial prediction \cite{casfarsmi}, and the  
interpretation of cost functions when the underlying model structure is 
unknown. 

This work was supported by EC grant ERBFMBICT950203, ONR grant 
N00014-99-1-0056, and Pembroke College, Oxford.

\begin{figure}
\caption{Least squares estimate of $a$ as a function of 
noise level using the analytic result (\ref{e:aopt}) corresponding to 
an infinite data set. 
The underlying system is the Logistic map with $a_0$ = 2. 
Parameter $a$ is systematically underestimated for both normally 
distributed noise (solid), and uniformly distributed noise (dashed). 
Both deviate significantly from the correct value
(dot-dashed).}
\label{fig1}
\end{figure}

\begin{figure}
\caption{Logistic map: (a) bifurcation diagram illustrating the variation
in $\mu(x,a)$, (b) distribution of estimates contrasting 
the LS,TLS, and ML cost functions from 1000 realisations where 
$a_0$ = 1.85, $N$ = 100, with normally distributed noise. 
The shading reflects the 95\% limits, the solid line the mean.}
\label{fig2}
\end{figure}

\begin{figure}
\caption{Moran-Ricker map:  
a comparison of LS,TLS, and ML cost functions 
for 200 realisations with $a_0$ = 3.7, $N$ = 100 with normally 
distributed measurement errors. 
The shading is as in Fig. \ref{fig2}.}
\label{fig3}
\end{figure}

\begin{figure}
\caption{Value of cost function in parameter space for 
a 2D delay reconstruction of the H\'{e}non map for $a_0$ = 1.4, 
$b_0$ = 0.3, $N$ = 500, and a noise level of 0.05:  
(a) $C_{\rm LS}$ and (b) $C_{\rm ML}$.}
\label{fig4}
\end{figure}

\end{document}